# Ultrafast creation of overlapping Rydberg electrons in an atomic BEC and Mott-insulator lattice

M. Mizoguchi[1,2], Y. Zhang[1,3], M. Kunimi[1], A. Tanaka[1], S. Takeda[1,2]†, N. Takei[1,2]‡, V. Bharti[1], K. Koyasu[1,2], T. Kishimoto[4], D. Jaksch[5,6], A. Glaetzle[5,6], M. Kiffner[5,6], G. Masella[7], G. Pupillo[7], M. Weidemüller[8,9], and K. Ohmori[1,2]*

[1]*Institute for Molecular Science, National Institutes of Natural Sciences, 38 Nishigo-Naka, Myodaiji, Okazaki, Aichi 444-8585, Japan.*
[2]*SOKENDAI (The Graduate University for Advanced Studies), Myodaiji, Okazaki, Aichi 444-8585, Japan.*
[3]*State Key Laboratory of Quantum Optics and Quantum Optics Devices, Institute of Laser Spectroscopy, and Collaborative Innovation Center of Extreme Optics, Shanxi University, Taiyuan, Shanxi 030006, China.*
[4]*Department of Engineering Science and Institute for Advanced Science, University of Electro-Communications, 1-5-1 Chofugaoka, Chofu, Tokyo 182-8585, Japan.*
[5]*Clarendon Laboratory, University of Oxford, Parks Road, Oxford, OX1 3PU, United Kingdom.*
[6]*Center for Quantum Technologies, National University of Singapore, 3 Science Drive 2, Singapore, 117543, Singapore.*
[7]*icFRC and ISIS (UMR 7006), Université de Strasbourg and CNRS, 67000 Strasbourg, France.*
[8]*Physikalisches Institut, Universität Heidelberg, Im Neuenheimer Feld 226, 69120 Heidelberg, Germany.*
[9]*Hefei National Laboratory for Physical Sciences at the Microscale and Department of Modern Physics, University of Science and Technology of China, Hefei, Anhui 230026, China, and CAS Center for Excellence and Synergetic Innovation Center in Quantum Information and Quantum Physics, University of Science and Technology of China, Shanghai 201315, China.*
*Corresponding author. E-mail: ohmori@ims.ac.jp
(Dated: October 11, 2019)

An array of ultracold atoms in an optical lattice (Mott insulator) excited to a state where single electron wave-functions spatially overlap would represent a new and ideal platform to simulate exotic electronic many-body phenomena in the condensed phase. However, this highly excited non-equilibrium system is expected to be so short-lived that it has eluded observation so far. Here, we demonstrate the first step toward its realization by exciting high-lying electronic (Rydberg) states of the atomic Mott insulator with a coherent ultrashort laser pulse. Beyond a threshold principal quantum number where Rydberg orbitals of neighboring lattice sites overlap with each other, the atoms efficiently undergo spontaneous Penning ionization resulting in a drastic change of ion-counting statistics, sharp increase of avalanche ionization and the formation of an ultracold plasma. These observations signal the actual creation of exotic electronic states with overlapping wave functions, which is further confirmed by a significant difference in ionization dynamics between a Bose-Einstein condensate and a Mott insulator.

Electronic properties of condensed matter are often determined by an intricate competition between kinetic energy, that aims to overlap and delocalize electronic wave functions across the crystal lattice, and localizing electron-electron interactions. In contrast, the gaseous phase is characterized by valence electrons being tightly localized around the ionic atom cores in discrete quantum states with well-defined energies. As an exotic hybrid of both cases, one may wonder which state of matter is created when a gas of isolated atoms is suddenly excited to a state where electronic wave functions spatially overlap as in the solid. How fast can those overlapping wave functions be created, and into which state do they finally disintegrate? If the timescales for these processes are sufficiently separated, one could use such a system as a quantum simulator for exotic many-body electronic phenomena dominated by the Coulomb interaction.

Here, we have explored first steps into such a regime by exciting high-lying electronic (Rydberg) states in an atomic Bose-Einstein condensate (BEC) and a unity-filling Mott insulator (MI), i.e. an optical lattice with one atom per site, with a coherent ultrashort laser pulse [1]. Beyond a threshold principal quantum number, we enter the regime of overlapping Rydberg electrons [2-5] that leads to a sharp increase of avalanche ionization of the gas and the formation of a plasma after dwell times of tens of nanoseconds. Understanding the dynamics of this process using full ion-counting statistics represents a major step toward the controlled creation of delocalized electronic many-body states.

An ensemble of Rydberg atoms in an optical lattice provides an ideal platform for the study of strongly interacting many-body systems [6-8]. Optical lattices have already been effective in studying Rydberg blockade [9-11], Rydberg crystals [12, 13], quantum simulation of Ising Hamiltonians [14, 15], bistability in open many-body systems [16], Rydberg molecules [17], atom-clock networks [18] and implementation of quantum random walks [19]. In these studies, continuous-wave excitation is employed to achieve precise quantum state control. Spectral resolution is typically much finer than the scale of the Rydberg-Rydberg interactions thus singling out specific few- or many-body quantum states of the structured gas involving single (Rydberg blockade [20]) or multiple (Rydberg molecules [21]) excitations. In the current experiment we explore the opposite regime using broadband excitation with picosecond laser pulses. As the spectral width covers the entire range of Rydberg interaction energies, the laser field might, in principle, coherently excite delocalized electronic superposition states extending over multiple atoms.

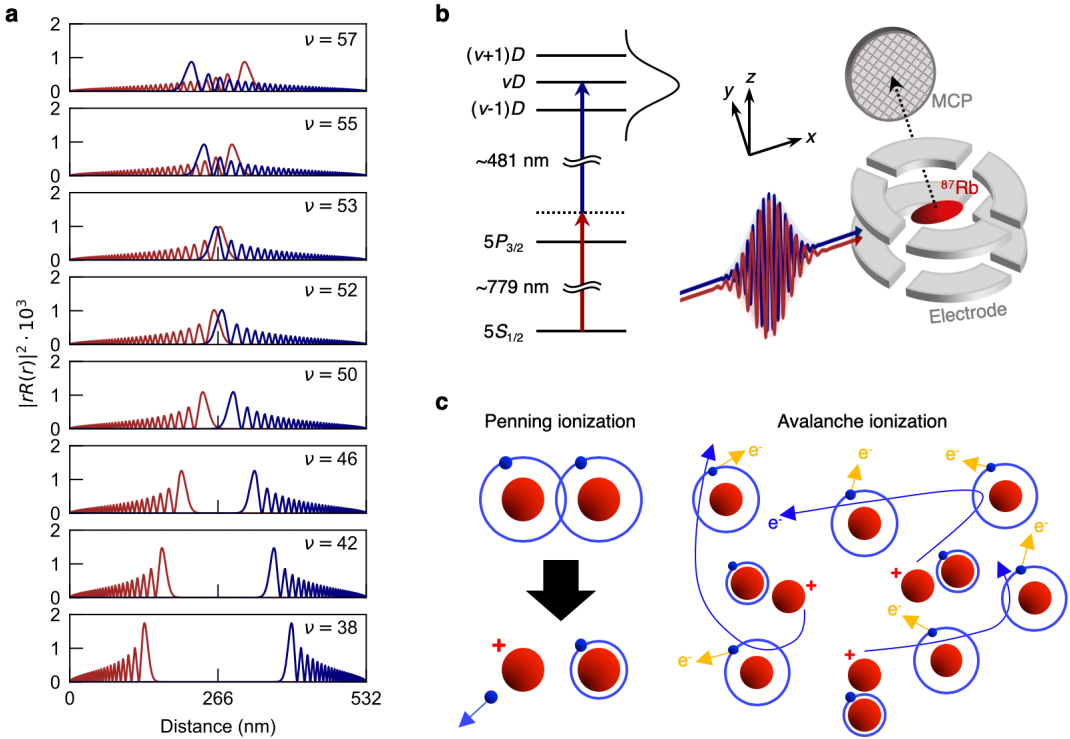

FIG. 1: **Schematic of the experiment. a**, Calculated Rydberg orbitals of the neighboring lattice sites for principal quantum numbers $\nu$ = 38, 42, 46, 50, 52, 53, 55 and 57 (from the bottom to the top), where the lattice spacing is 532 nm. **b**, $^{87}$Rb atoms in a Bose-Einstein condensate or a unity-filling Mott insulator are excited to Rydberg D states via a two-photon transition using picosecond laser pulses. Ions created through ionization processes are accelerated by the electric field pulse applied to electrodes and subsequently accelerated to a microchannel plate (MCP) detector. **c**, Schematic presentation of the two-step mechanism of Rydberg-induced plasma formation. The first step is Penning ionization from a pair of Rydberg atoms with overlapping electron orbitals. These Penning electrons (blue arrows) may further ionize other Rydberg atoms (yellow arrows) in an avalanche process. The electrons might eventually be trapped by the remaining ion cores, thus forming an ultracold plasma.

Under our experimental conditions, the electronic Rydberg wave functions of a pair of nearest neighbors in the lattice overlap for principal quantum numbers beyond $\nu \geq 50$ (see Fig. 1a). We can therefore study the effects of overlapping electronic states on the Rydberg gas selectively and with high temporal resolution. Employing full-counting statistics of ion detection under such extremely controlled conditions, we have found that the characteristics of the subsequent ionization processes is dramatically changed by the degree of initial overlap of Rydberg orbitals between neighboring atoms. In particular, the formation of an ultracold plasma is fully suppressed if the initial Rydberg pair wave functions are not overlapping.

Figure 1b schematically shows our experimental system. The experiment starts by preparing a Bose-Einstein condensate (BEC) of about $3 \times 10^4$ $^{87}$Rb atoms which is then adiabatically loaded into an optical lattice to produce a unity-filling Mott insulator of about $3 \times 10^4$ $^{87}$Rb atoms in the hyperfine state $|F = 1, m_F = 1\rangle$ or $|F = 2, m_F = 2\rangle$ of the $5S_{1/2}$ ground state manifold (see the subsection "Ultracold atom preparation" in the Supplementary Materials [22]). In the Mott insulator, the atoms form a three-dimensional lattice with a period of 532 nm. The $^{87}$Rb atoms are excited to Rydberg states via a two-photon transition using broadband picosecond laser pulses with center wavelengths tuned to ∼ 779 nm (IR pulse) and ∼ 481 nm (blue pulse), as schematically shown in Fig. 1b, after turning off the trapping laser. The bandwidth of the excitation is ∼ 140 GHz. When the atoms are initially prepared in the $|F = 2, m_F = 2\rangle$ state, the laser pulses are circularly polarized along the direction of the magnetic field, effectively suppressing excitations to the Rydberg S states. Accordingly, the state $\nu D_{5/2}$ ($m_J = 5/2$) is mostly populated, with the principal quantum number $\nu$ selected by tuning the center wavelength of the blue pulse. When the atoms are initially populated in the $|F = 1, \ m_F = 1\rangle$ state, the direction of the magnetic field is set perpendicular to the propagation axis of the laser pulses so essentially the state $\nu D_{5/2}$ is populated with a ~ 2% admixture of the state $\nu D_{3/2}$ (see [22] for more details). For $\nu < 47$, almost a single Rydberg state is excited, whereas beyond this principal quantum number the excitation bandwidth exceeds the level spacing, and a Rydberg wave packet of two or more Rydberg levels is created. The Rydberg excitation probability $p$ is controlled by tuning the blue pulse energy and measured by field ionization with an electric field pulse (rise time ~ 500 ns, fall time ~ 300 ns) combined with a microchannel plate (MCP) detector (see Fig. 1b and the subsections "Electrodes and the electric field pulse" and "Excitation probability measurements" in [22]).

As schematically shown in Fig. 1c, it is well established that there are generally two processes leading to the ionization of interacting Rydberg atoms and, under certain circumstances, the formation of an ultracold plasma [23, 24]. The initial ionization process following electronic excitation is Penning ionization of a pair of interacting Rydberg atoms [4, 5, 25, 26]. In a secondary process, the free electrons might ionize other Rydberg atoms. This avalanche process is particularly efficient when the accumulated Penning ions trap the electrons, thus leading to the formation of an ultracold plasma [27, 28]. To detect the ions created via these ionization processes, we accelerate those ions toward the MCP detector with the electric field pulse, whose voltage is lowered enough not to field-ionize Rydberg states $\nu < 63$ (see the subsection "Ion counting" in [22]). The delay $\tau$ between the picosecond laser pulses and the rising edge of the electric field pulse is set to ~ 60 ns, thus ensuring no temporal overlap between laser and electric field pulses (see [22] for more details of the definition of $\tau$). The experiment is repeated many times under identical experimental conditions, allowing for a statistical analysis of the number of ions being produced after each realization.

In a first series of experiments we compare ion-counting statistics for a BEC and a MI under similar conditions (initial atom number ~ 30,000, volume and Rydberg excitation probability $p \sim 0.03$) exciting atoms to the state $\nu = 42$. Thus, the major difference between these two scenarios is the spatial pair correlation function, which excludes pairs at distances below the lattice spacing of 532 nm in the case of the MI, while permitting any pair distance in the case of the BEC (for studies on Rydberg excitation of a BEC in the blockade regime, see [29]). Figure 2a compares the arrival-time distribution of ion signals at the MCP detector for the BEC (upper panel) and MI (lower panel). A first burst of ions arrives around 45 μs given by the time of flight after being accelerated by the electric field pulse. These ions are created in the 60-ns dwell time between the laser and electric field pulses. There is a tail of ions arriving at later times, more strongly pronounced in the MI case than for the BEC, which can be assigned to ions created during the ~ 800-ns electric field pulse, thus experiencing a smaller average acceleration. As these latter ions are created under conditions not well defined, we ignore ions arriving at times later than or equal to 60 μs in the analysis throughout this paper.

The most striking feature of the comparison between the BEC and MI scenarios shown in Fig. 2a is the drastic suppression of ionization, indicating a vanishing probability for Penning ionization during the 60 ns period of free evolution in the MI case. In fact, efficient Penning ionization would require substantial overlap of the electronic wave functions [4,5], which is excluded in the MI for the present principal quantum number $\nu = 42$ (see Fig. 1a). Further insight into the dynamics following initial Penning ionization is provided by inspecting the statistical distribution of ion numbers, in particular when normalizing it to the average number of initial Rydberg atoms, as depicted in Fig. 2b. While a negligible portion of Rydberg atoms is actually ionized in the MI case, one third of the Rydberg atoms excited in the BEC get ionized on average, with a large spread ranging from few realizations without any ionization to ionization of more than half of initial Rydberg number.

In order to gain further insights into the underlying dynamics of ion formation, we study the spontaneous ionization after Rydberg excitation in a MI in more detail, employing the same experimental conditions (initial atom number ~ 30,000, volume and Rydberg excitation probability $p \sim 0.03$) as before, but now turning off a magnetic field gradient to support the BEC against the gravity and exciting from the $|F = 2, \ m_F = 2\rangle$ state instead of the $|F = 1, \ m_F = 1\rangle$ state to populate the almost single magnetic sublevel $\nu D_{5/2}$ ($m_J = 5/2$). Figure 3a shows the full-counting statistics of the ion number for principal quantum numbers $\nu$ ranging from 38 (no wave function overlap of neighboring Rydberg pairs, see Fig. 1a) to 57 (substantial wave function overlap). The distribution of ion numbers changes drastically around principal quantum number $\nu \sim 50$. While only a small portion of Rydberg atoms (around 10% on average) is ionized for lower $\nu$, a much larger fraction becomes ionized at higher $\nu$ with a non-negligible fraction of realization where more than 2/3 of the initial atoms are ionized.

A more quantitative analysis of the ion distribution statistics is provided by Figs. 3b-d, which depict the median normalized to the average number of initial Rydberg atoms, the Q parameter as a measure of the deviation from a Poissonian distribution (Q = 0), and the fraction of realizations resulting in the ionization of more than 2/3 of the initial Rydberg atoms, respectively. All three measures exhibit a sharp change in their dependence on the principal quantum number separating the regime $\nu < 50$ from $\nu > 50$.

In order to interpret the strong super-Poissonian statistics (Q > 0) for $\nu > 50$ as captured by the Q parameter defined as Q = $\langle (N - \langle N \rangle)^2 \rangle / \langle N \rangle - 1$ (see Fig. 3c), where $N$ is the number of ions produced in each measurement, we have numerically calculated the statistics of the Penning ionization, assuming that a pair of Rydberg atoms on neighboring lattice sites undergoes solely Penning ionization (see Fig. 1c, left panel, with very few examples where four or more atoms are arranged without a gap in the lattice sites), and found that the ion distribution would then follow Poissonian statistics (see [22] for the details), so that this possibility can be safely excluded. The observed super-Poissonian distribution thus points toward a different mechanism of ion creation, which we identify with secondary avalanche ionization following Penning ionization initiated through laser excitation of overlapping electronic wave functions of Rydberg pairs in the nearest-neighbor lattice sites (see Fig. 1c, right panel). The avalanche ionization constitutes a highly nonlinear process critically depending on the initial microscopic spatial configuration of Rydberg atoms in each experimental realization, thus giving large values for the Q parameters.

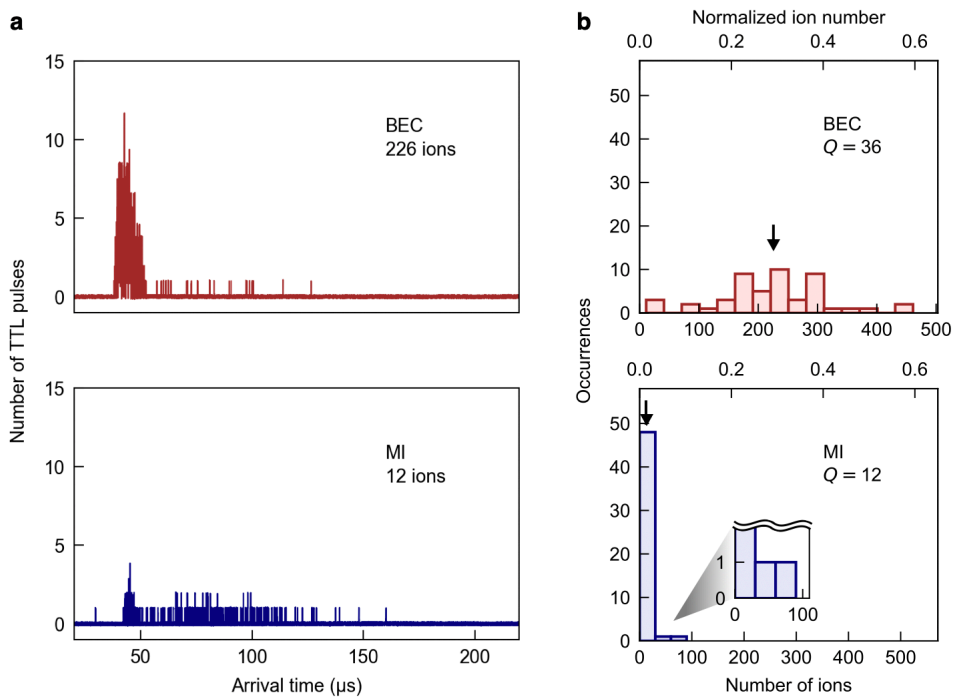

FIG. 2: **Comparison of ion productions from a Bose-Einstein condensate (BEC) and unity-filling Mott insulator (MI).** Atoms are excited to the Rydberg state (principal quantum number $\nu = 42$, excitation probability $p \sim 0.03$) with picosecond laser pulses from the $^{87}$Rb ground state $5S_{1/2} |F = 1, m_F = 1\rangle$, and produced ions are detected after a dwell time of $\tau \sim 60$ ns. The graphs show the results of fifty independent measurements under similar conditions ($\sim 30{,}000$ atoms, volume) **a**, Arrival-time distribution of the ion signals. The ordinate represents the number of TTL pulses of ion counting summed over the fifty measurements (see [22] for the details of ion counting). The Penning and avalanche ions produced during the dwell time are accelerated toward the MCP detector with the electric field pulse (pulse width $< 1$ μs), whose voltage is lowered enough not to ionize Rydberg states $\nu < 63$. The ion number indicated in each figure is the one detected at arrival times shorter than 60 μs and averaged over the fifty measurements. These numbers are rounded to an integer. **b**, Statistical distribution of ion numbers produced from the BEC (top graph) and MI (bottom graph). The bin size is 30 ions. The arrows indicate the mean number of ions. The upper abscissa shows the number of ions normalized to the average number of Rydberg atoms created initially. The Q parameter quantifying the deviation from a Poissonian distribution is 36 for the BEC and 12 for the MI, respectively.

Transformation of more than 2/3 of the Rydberg atoms into ions signals the emergence of an ultracold plasma [23, 26, 30]. Fig. 3d indicates that beyond $\nu \sim 50$, an increasing fraction of microscopic realizations of efficient avalanche ionization culminates in the creation of such an exotic neutral plasma, while plasma formation through short-pulse Rydberg excitation of a MI is not favored for $\nu < 50$. We have numerically evaluated the depth of the ionic potential formed by Penning ions under the present experimental conditions, assuming that a pair of Rydberg atoms in the neighboring lattice sites with their overlapping Rydberg orbitals undergoes Penning ionization (see [22] for the details). We have averaged calculated ionic potentials over many different initial configurations of Rydberg atoms randomly generated by the Monte Carlo method. It is found that the spatially averaged potential depth is indeed large enough to trap Penning electrons and could thus indeed result in efficient plasma formation. An intriguing feature of this ultracold plasma is the preordering of the remaining ion cores, which might result in reduced disorder induced heating and thus an increased plasma parameter [31-33].

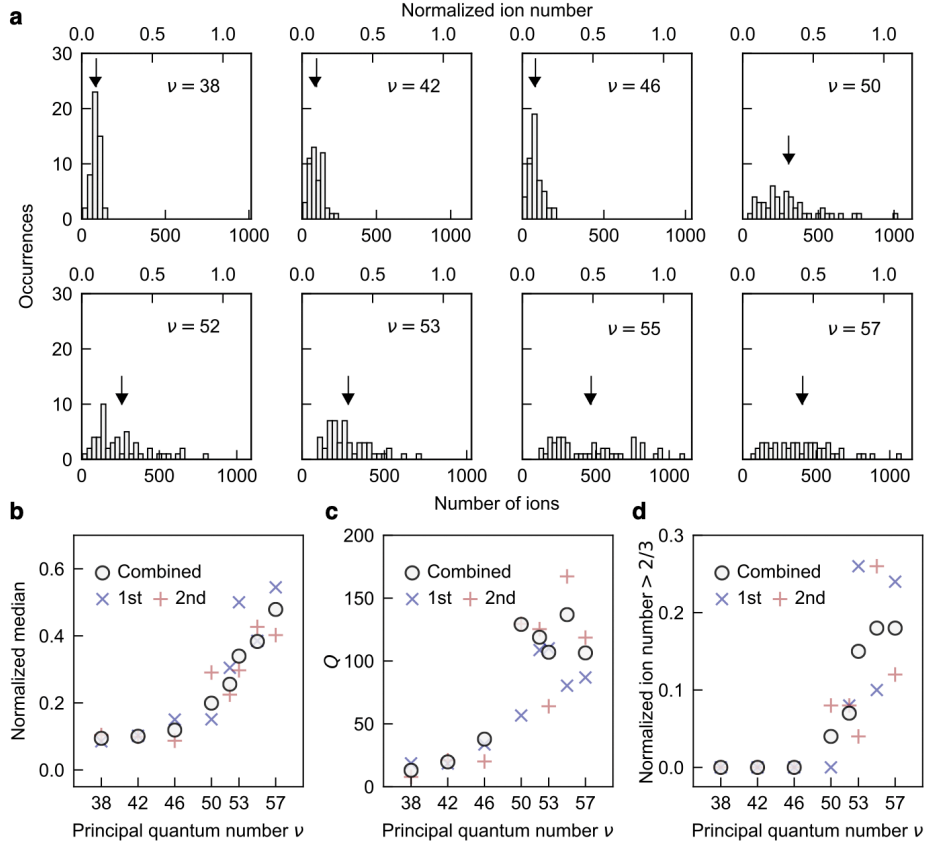

FIG. 3: **Statistics of ions produced after the ultrafast Rydberg excitation of a MI for different principal quantum number $\nu$.** Experimental conditions and graphical representation are identical to Fig. 2, except the initial state which is now the $5S_{1/2}$ $|F = 2, m_F = 2\rangle$ state. **a**, Statistical distribution of produced ion numbers for fifty independent measurements. The arrows denote the measured mean values. **b**, **c** and **d**, The median normalized to the average number of initial Rydberg atoms, Q parameter, and fraction of realizations where the normalized ion number is larger than 2/3 as a function of principal quantum number $\nu$. The blue crosses and red plus signs in **b**, **c** and **d** represent two data sets of fifty measurements made on different days, respectively. Those two data sets have been merged and represented by the black circles.

The results of this second series of experiments involving a unity-filling MI can be summarized in the following simple model shedding new light onto the ionization dynamics of an ultracold Rydberg gas [4, 5, 31, 34-36]: When Rydberg excitation probability is low and wave function overlap is negligible, corresponding to the regime $\nu < 50$, ionization occurs only through accidental primary ions being created, e.g., by motion-induced Penning ionization [37] or other processes. The electrons freed by this process might eventually ionize other Rydberg atoms in an avalanche-like process. Only a small fraction of the Rydberg gas gets ionized actually (or undergoes state redistribution [36]), and formation of an ultracold plasma is excluded. If, however, spontaneous initial Penning ionization is facilitated by laser excitation of overlapping Rydberg pair states through the broadband laser field, as observed in the regime $\nu > 50$ in the present experiment, subsequent avalanche processes result in ionization of large fractions of Rydberg atoms and efficient formation of an ultracold plasma even for low Rydberg excitation probabilities.

This reasoning based on the results involving a MI also explains the striking contrast between the BEC and MI scenarios shown for $\nu = 42$ in Fig. 2. In the BEC, the atoms are randomly distributed without restriction on the minimal pair distance, so that there are always pairs of atoms for which overlapping Rydberg orbitals are excited even for low principal quantum numbers. These pairs form the initial grains of Penning ionization triggering the highly nonlinear avalanche processes, whereas in the MI such pairs are absent. In contrast to the MI scenario at higher principal quantum number, however, ionization in the BEC for $\nu = 42$ does not necessarily lead into the formation of an ultracold plasma since there is no realization where more than 2/3 of the initial Rydberg atoms are ionized, as seen in Fig. 2b. Under the present experimental conditions the densities of Penning ions may not be large enough to maintain an effective trapping potential for the electrons. This is consistent with our numerical evaluation of the depth of the ionic potential formed by Penning ions, where the potential depth is much larger than expected kinetic energy of Penning electrons in the MI for $\nu > 50$, whereas they are comparable in the BEC for $\nu = 42$ under the present experimental conditions (see [22] for more details).

The precisely controlled ionization and plasma formation demonstrated in this work provides a novel path toward the study of the competition between kinetic energy and electron-electron interactions in crystal structures. This competition underlies a vast range of the most elusive phenomena in strongly correlated physics. The present work provides us with a better understanding of the stability of many-body Rydberg systems [14, 15, 38-41] and the mechanism for ultracold plasma formation [23, 24, 28, 42]. Importantly, on timescales much shorter than the 60 nanoseconds dwell time observed in our experiments coherent electron dynamics can in principle generate metal-like phases in which Rydberg electrons are shared by multiple sites of a Mott insulator [1]. For a pair of atoms with overlapping electronic wave packets the time scale for Penning ionization is around 1 - 10 nanoseconds for $\nu \gtrsim 50$ studied here [4, 5]. This allows observation of delocalized electron states using ultrafast Ramsey interferometry with attosecond precision on picosecond timescales [43-45] before they decay through Penning ionization. Such a metal-like phase would open a completely new regime of many-body physics that could be applied to quantum simulation of exotic many-body electronic phenomena.

**Acknowledgments:** We are grateful for insightful discussions with Tom Killian, Ed Grant, Francis Robicheaux and Kaden Hazzard. We also thank Hisashi Chiba and Yasuaki Okano for technical support. This work was supported by Japan's quantum-technology flagship program "Q-LEAP" by MEXT and JSPS Grant-in-Aid for Specially Promoted Research Grant Number 16H06289. D.J. acknowledges support by EPSRC grant No. EP/P009565/1 and the European Research Council under the European Union's Seventh Framework Programme (FP7/2007-2013)/ERC Grant Agreement No. 319286 Q-MAC. M.W. acknowledges partial support by the Heidelberg Center for Quantum Dynamics, by the DFG Priority Program "GiRyd 1929" (DFG WE2661/12-1), the DFG Collaborative Research Center "SFB 1225 (ISOQUANT)", and the European Union H2020 FET Quantum Flagship project PASQuanS (Grant No. 817482). K. O. thanks Alexander von Humboldt foundation, University of Heidelberg and University of Strasbourg for supporting this international collaboration.

---

†Present address: Department of Applied Physics, School of Engineering, The University of Tokyo, 7-3-1 Hongo, Bunkyo-ku, Tokyo 113-8656, Japan.
‡Present address: Department of Physics, Graduate School of Science, Kyoto University, Kyoto 606-8502, Japan.

# Supplementary Materials for "Ultrafast creation of overlapping Rydberg electrons in an atomic BEC and Mott-insulator lattice"

## Ultracold atom preparation

A 3D magneto-optical trap (MOT) of $^{87}$Rb atoms was loaded from a 2D-MOT for 2 s. During the subsequent MOT compression for 50 ms, atoms were transferred to a crossed optical dipole trap (ODT) formed by 1,064 nm beams. A Bose-Einstein condensate (BEC) of $\sim 3 \times 10^4$ atoms in the $5S_{1/2}$, $|F=1, m_F=1\rangle$ state was created after the evaporative cooling in the ODT for 8.05 s. During the next 100 ms, the potential depth was decreased and a magnetic field gradient of $\sim 30$ G/cm was turned on to cancel out the gravity for this state. We then loaded the BEC into a three-dimensional optical lattice with a lattice spacing of 532 nm for the next 300 ms to form a unity-filling Mott insulator (MI). The final lattice depth was 20 $E_r$, where $E_r$ is the recoil energy of the optical lattice beam. For the results shown in Fig. 2 in the main text, both the BEC and unity-filling MI were held in the trapping potential having almost same external trapping frequencies $(\omega_x, \omega_y, \omega_z) = 2\pi \times (20, 47, 47)$ Hz (see Fig. 1b in the main text). For the results shown in Fig. 3 in the main text, we turned off the magnetic field gradient after the lattice ramp and applied a microwave pulse to perform an adiabatic rapid passage from $|F=1, m_F=1\rangle$ to $|F=2, m_F=2\rangle$ state.

## Electrodes and the electric field pulse

Figure S1 shows the schematic of electrodes and a microchannel plate (MCP) to detect Rb$^+$ ions. Eight electrodes were fixed inside a vacuum chamber to surround an atomic ensemble. We applied positive and negative voltage pulses to red and blue electrodes, respectively. For the results shown in Fig. 2 in the main text, we applied only negative voltage pulses whose voltage was set to -150 V. For the results shown in Fig. 3 in the main text, we applied both positive and negative voltage pulses whose voltages were set to 100 V and -50 V, respectively. The resulting electric field was $\sim 20$ V/cm. When we measured the Rydberg excitation probability, we applied both positive and negative voltage pulses whose voltages were set to 1500 V and -750 V, respectively. The rise and fall times of all the voltage pulses were $\sim 500$ ns and $\sim 300$ ns, respectively.

## Excitation probability measurements

In order to estimate the Rydberg excitation probability of a unity-filling MI, we repeated the Rydberg excitation and field ionization with high voltage pulses (see the subsection "Electrodes and the electric field pulse") for 200 times at a 1 kHz repetition rate. We integrated the obtained ion signals using a gated integrator. The integrated value $y$ followed the function $y = A(1-p)^{x-1}$, where $A$ is a factor proportional to the number of Rydberg atoms initially prepared, $p$ is the Rydberg excitation probability and $x$ is the number of irradiated pulses ranging from 1 to 200. We estimated the excitation probability $p$ by fitting the measured decay with this function. We also measured the relation between the factor $A$ and the excitation probability $p$ under the condition of ~ 30,000 atoms in a unity-filling MI, which was used to estimate the excitation probability of a BEC. For the case of the BEC, we found that the ground state atoms escaped from the trap even if we used only IR pulses. This could be due to the shallower trapping potential for the BEC than MI. Accordingly, we shined only one excitation pulse to measure the factor $A$ after preparing the BEC of ~ 30,000 atoms. We then converted this factor to the excitation probability using the relation between $A$ and $p$ mentioned above. We controlled the excitation probability $p$ by tuning the blue pulse energy and kept the IR pulse energy always the same.

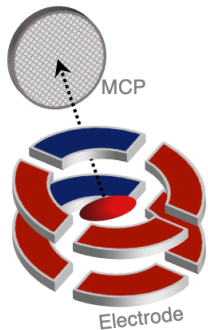

FIG. S1: **Schematic of electrodes and a microchannel plate.** Eight electrodes are fixed inside a vacuum chamber. We apply positive and negative voltage pulses to red and blue electrodes, respectively to detect ions with a microchannel plate.

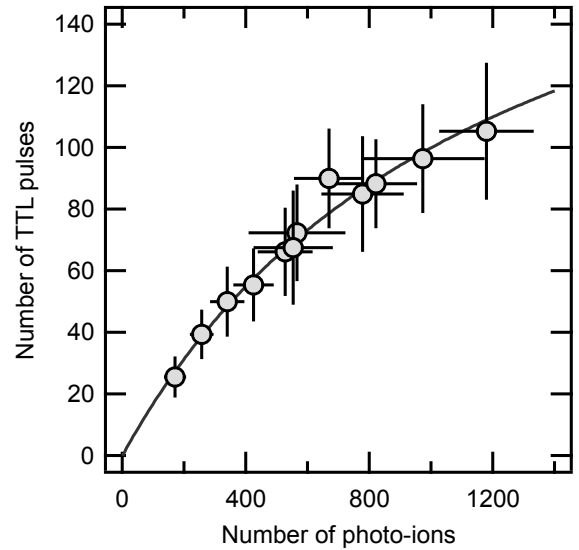

FIG. S2: **Calibration curve to estimate number of produced ions.** The relation between the number of TTL pulses and photo-ions are obtained from the calibration measurements. The black circles represent the measured data and solid curve represents the result of fitting with the function $N_{\text{TTL}} = \eta N_{\text{ion}}/(1 + \tau N_{\text{ion}})$. The error bars represent the standard deviation.

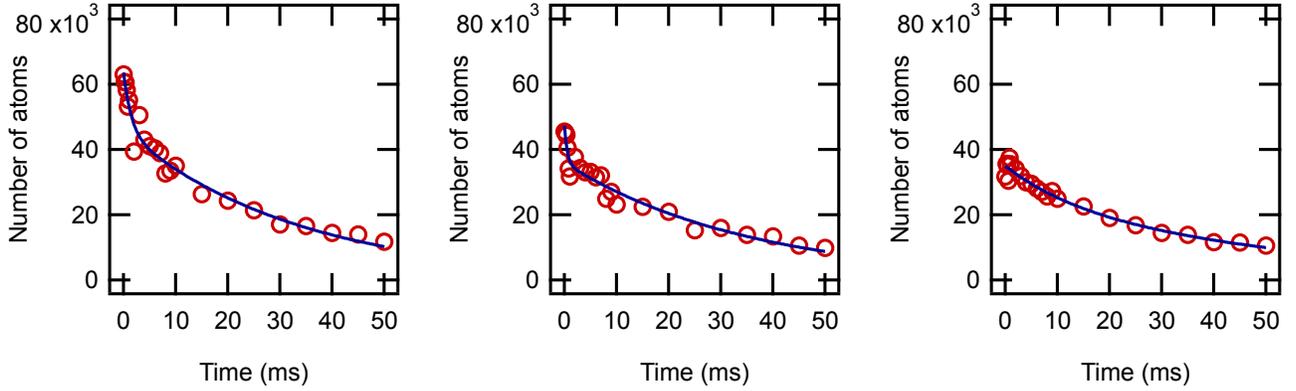

FIG. S3: **Light-assisted collisions in a Mott insulator.** The red circles represent the measured data and the blue lines represent the fitting results with the sum of two exponential functions. Initial atom numbers are 63,000, 46,000 and 32,000 (from left to light).

### Ion counting

In order to detect the ions created via the ionization processes discussed in the main text, we lowered the voltage of the electric field pulse (see the subsection "Electrodes and the electric field pulse") enough not to ionize Rydberg states $\nu < 63$. The output of the MCP was amplified with a preamplifier and sent to a discriminator which generates TTL pulses. In order to obtain the relation between the number of TTL pulses and produced ions, we carried out a calibration measurement consisting of three independent measurements by using photo-ions. First, we detected photo-ions and counted the number of TTL pulses generated due to these ions. Here, we used only IR pulses and created photo-ions by a three-photon ionization of a MI. The volume of the MI is almost same as the one used in Figs. 2 and 3 in the main text. Second, we measured the photo-ionization probability employing the same method used to estimate the Rydberg excitation probability (see the subsection "Excitation probability measurements"). Third, we measured the atom number by the absorption imaging. Number of photo-ions were obtained as a product of the photo-ionization probability and atom number. We repeated these measurements by changing the photo-ionization probability and/or atom number to obtain the relation between the number of TTL pulses and photo-ions. Figure S2 shows the results of this calibration measurement. We then obtained the calibration curve by fitting the data based on the non-extended dead-time model [46]. We used the fitting function $N_{\text{TTL}} = \eta N_{\text{ion}}/(1 + \tau N_{\text{ion}})$, where $N_{\text{TTL}}$ and $N_{\text{ion}}$ is number of TTL pulses and photo-ions, respectively. The fitting parameters $\eta$ and $\tau$ was estimated to be 0.18307 and 0.00083146, corresponding to the detection efficiency and dead time, respectively. We used this function $N_{\text{TTL}} = \eta N_{\text{ion}}/(1 + \tau N_{\text{ion}})$ to estimate the produced ion number from the measured TTL pulse number in Figs. 2 and 3 in the main text.

### Verification of unity-filling Mott insulator

We made an independent measurement of the light-assisted collisions [47] in a MI as a function of atom number with the geometrical mean trapping frequency of 39 Hz. Figure S3 shows the result, from which it is seen that the fast decay component (corresponding to doublons) disappears for atom number less than ~ 32,000. The fraction of doublons is thus almost negligible under the present conditions (atom number ~ 30,000 and the geometrical mean trapping frequency 35 Hz).

### Fraction of Rydberg populations

The picosecond laser pulses are circularly polarized along the x-axis (see Fig. 1b in the main text). When atoms are initially prepared in the $5S_{1/2}, |F = 1, m_F = 1\rangle$ state, the direction of the magnetic field during the Rydberg excitation is along the z-axis perpendicular to the propagation axis of the laser pulses. Therefore, not only the state $\nu D_{5/2}$ ($m_J = 5/2$) but also other states are populated after the Rydberg excitation. Figure S4 shows the calculated fraction of their populations using the open source program [48]. It is seen from this figure that the state $\nu D_{5/2}$ is mostly populated with a ~ 2% admixture of the state $\nu D_{3/2}$.

| | $m_J =$ −5/2 | −3/2 | −1/2 | +1/2 | +3/2 | +5/2 |
|---|---|---|---|---|---|---|
| $\nu D_{5/2}$ | 7.7% | 25.1% | 31.7% | 21.5% | 9.7% | 2.6% |

| | $m_J =$ | −3/2 | −1/2 | +1/2 | +3/2 | |
|---|---|---|---|---|---|---|
| $\nu D_{3/2}$ | | 0.2% | 0.7% | 0.7% | 0.2% | |

FIG. S4: **Calculated fraction of Rydberg populations.** When atoms are initially prepared in the $|F = 1, m_F = 1\rangle$ state, the state $\nu D_{5/2}$ and $\nu D_{3/2}$ are populated after the Rydberg excitation.

### Definition of the delay $\tau$

The electric field pulse consists of positive and negative voltage pulses (see the subsection "Electrodes and the electric field pulse"). The delay $\tau$ between the picosecond laser pulse and the rising edge of the electric field pulse is set to ~ 60 ns, thus ensuring no temporal overlap between laser and electric field pulses. Figure S5 shows their timing relationship. Here the positive and negative voltages are set to 100 V and -50 V, respectively. The laser pulse is monitored by a photo-detector. The delay $\tau$ is defined as the delay between the rising edge of the monitored laser pulse and electric field pulse.

### Monte Carlo simulation for Penning ionization

Here, we explain how to perform the Monte Carlo simulation for the ion-counting statistics due to the Penning ionization in the MI and BEC cases. In the MI case, we consider a three-dimensional cubic lattice $(N_x, N_y, N_z) = (56, 23, 23)$, where $N_{x,y,z}$ is the number of lattice sites for each direction. We assume that a Rydberg atom occupies a lattice site with probability $p$. In order to consider the Penning ionization, we define a cluster, which is a group of nearest neighboring occupied sites. A cluster size is defined by the number of Rydberg atoms containing the cluster.

We assume that when two Rydberg atoms are next to each other, one ion is created due to the Penning ionization. When the cluster size is larger than two, we assume that $\lfloor m/2 \rfloor$ ions are created, where $m$ is the cluster size and $\lfloor \cdot \rfloor$ is the floor function. For example, the size 3 cluster generates one ion and the size 4 cluster generates two ions, respectively.

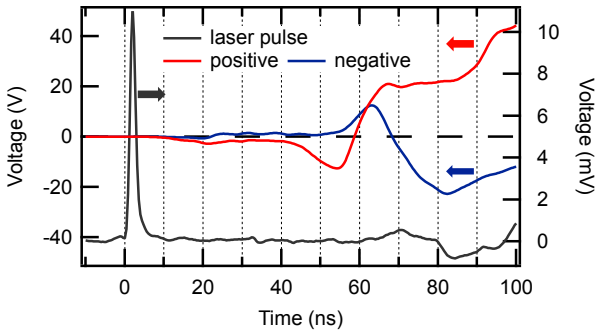

FIG. S5: **Timing relationship between the laser and electric field pulses.** The black line represents the laser pulse monitored by a photo-detector. The red and blue lines represent the positive and negative voltage pulses, respectively (see the subsection "Electrodes and the electric field pulse").

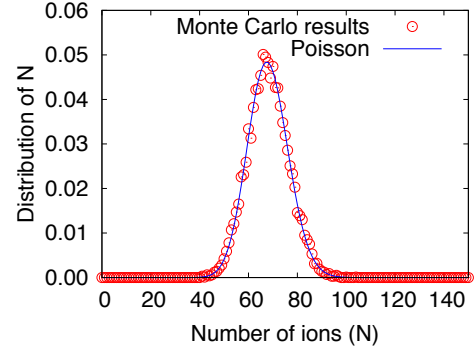

FIG. S6: **Monte Carlo results of the ion number distribution of the MI for $p = 0.03$.** The red circles represent the Monte Carlo results and blue curve represents the Poisson distribution function for the mean number of ions $N_{\text{mean}} \simeq 68$ obtained by the Monte Carlo simulations.

Figure S6 shows the Monte Carlo results of the ion number distribution for $p = 0.03$ in the MI case. The results are obtained by 10,000 samples. The red points represent the Monte Carlo results and the blue line represents the Poisson distribution function. We plot the Poisson distribution function by using the mean number of ions $N_{\text{mean}} \simeq 68$ estimated by the Monte Carlo method. We can see that the Monte Carlo results and the Poisson distribution function are in good agreement. We also evaluate the Q parameter by using the Monte Carlo method. The value is about 0.02. From these results, we can conclude that the ion number distribution due to the Penning ionization for the MI follows a Poissonian.

In the BEC case, we consider the 30,000 $^{87}$Rb atoms in the harmonic trap, whose trapping frequencies are matching the experimental condition, i.e. $(\omega_x, \omega_y, \omega_z) = 2\pi \times (19.6, 46.8, 46.8)$ Hz. We randomly distribute 30,000 $^{87}$Rb atoms according to the Thomas-Fermi distribution function and then replace $^{87}$Rb atoms with the Rydberg atoms with probability $p$. To define the cluster for the BEC case, we consider the radius of the Rydberg atom $R_\nu$, where $\nu$ is a principal quantum number of the Rydberg atom. We define the radius of the Rydberg atom as the distance between the nucleus and the point where the amplitude of the wave function of the Rydberg electron is maximum (see Fig. 1a in the main text). The wave function of the Rydberg electron is obtained by using the open source program [48]. A cluster of the BEC is defined by a group of the Rydberg atoms whose inter-particle distance is less than or equal to $2R_\nu$. In the same manner of the MI case, we assume that $\lfloor m/2 \rfloor$ ions are generated for each size $m$ cluster.

Figure S7 shows the results for the ion number distribution for $p = 0.03$ for the BEC case. The results are obtained by 10,000

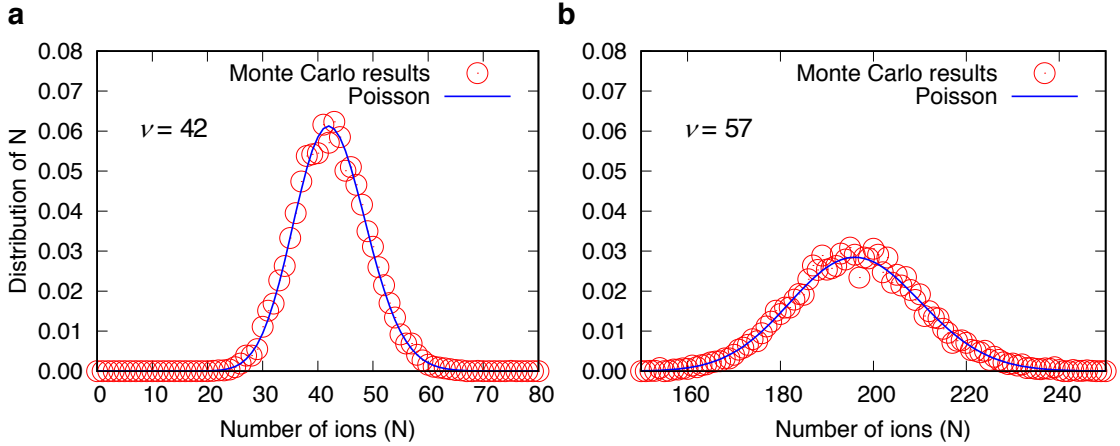

FIG. S7: **Monte Carlo results of the ion number distribution of the BEC for $p = 0.03$. a**, The ion number distribution for $\nu = 42$. **b**, The ion number distribution for $\nu = 57$. The red circles represent the Monte Carlo results and blue curve represents the Poisson distribution function for the mean number of ions $N_{\text{mean}} \simeq 42.5$ for $\nu = 42$ and $N_{\text{mean}} \simeq 197$ for $\nu = 57$, respectively.

samples. The red points represent the Monte Carlo results and the blue line represents the Poisson distribution function. We plot the Poisson distribution function by using the mean number of ions $N_{\text{mean}} \simeq 42.5$ for $\nu = 42$ (Fig. S7a) and $N_{\text{mean}} \simeq 197$ for $\nu = 57$ (Fig. S7b) estimated by the Monte Carlo method. The Q parameters are 0.006 for $\nu = 42$ and -0.03 for $\nu = 57$, respectively. These results show that the ion number distribution due to the Penning ionization for the BEC is also described by a Poissonian.

### Coulomb potential generated by ions

Here, we discuss the Coulomb potential created by ions [23]. Let $n_i(\boldsymbol{r})$ be an ion number density. An electron feels the following Coulomb potential:

$$U(\boldsymbol{r}) \equiv -\frac{e^2}{4\pi\varepsilon_0} \int d\boldsymbol{r}' \frac{n_i(\boldsymbol{r}')}{|\boldsymbol{r}-\boldsymbol{r}'|}, \quad (S1)$$

where $\varepsilon_0$ is the electric constant and $e$ is the elemental charge. We assume the ion number density as a summation of the Gaussian

$$n_i(\boldsymbol{r}) \equiv \sum_{j=1}^{N_I} \frac{1}{\sqrt{(2\pi\sigma^2)^3}} e^{-|\boldsymbol{r}-\boldsymbol{r}_j|^2/(2\sigma^2)}, \quad (S2)$$

where $N_I$ and $\boldsymbol{r}_j$ are the total number of the ions and the positions of the ions, respectively. The potential is given by

$$U(\boldsymbol{r}) = -\frac{e^2}{4\pi\varepsilon_0} \sum_{j=1}^{N_I} \frac{1}{|\boldsymbol{r}-\boldsymbol{r}_j|} \text{erf}\left(\frac{|\boldsymbol{r}-\boldsymbol{r}_j|}{\sqrt{2}\sigma}\right), \quad (S3)$$

where erf($\cdot$) is the error function. We choose $\sigma = 0.1$ nm. We have checked that the results presented below are insensitive to the choice of $\sigma$ around this value.

Here, we consider the spatial average of the potential. The averaged potential is defined by

$$U_{\text{ave}} \equiv \frac{1}{V} \int d\boldsymbol{r}\ U(\boldsymbol{r}), \quad (S4)$$

where $V$ is the volume of the integral region. For the MI case, we set the integral region as $[0, L_x] \times [0, L_y] \times [0, L_z]$, where $L_x = 30$ μm and $L_y = L_z = 13$ μm. The position of the $j$-th ion is given by $\boldsymbol{r}_j = d(n_{x,j}\boldsymbol{e}_x + n_{y,j}\boldsymbol{e}_y + n_{z,j}\boldsymbol{e}_z)$, where $\boldsymbol{e}_{x,y,z}$ is the unit vector of x, y, and z direction, respectively, $d = 532$ nm is the lattice constant, and $n_{x,j} = 1, 2, \cdots, 56$ and $n_{y,j}, n_{z,j} =$ 1,2,⋯,23. For the BEC case, we set the integral region as the region where the Thomas-Fermi density distribution function is nonzero. The volume of the integral region is given by $V = 4\pi R_{TF,x} R_{TF,y} R_{TF,z}/3$, where $R_{TF,x,y,z}$ is the Thomas-Fermi radius of each direction.

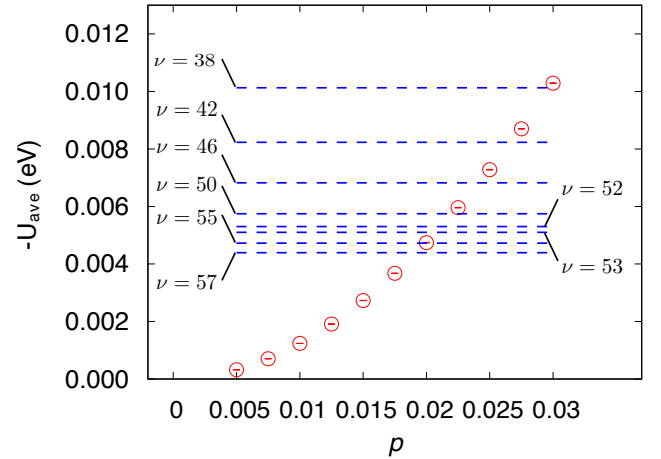

FIG. S8: **Monte Carlo results of the averaged potential created by the ions of the MI as a function of the Rydberg excitation probability $p$.** The red points represent the averaged potential and the blue dashed lines represent the binding energy for each principal quantum number. These results are obtained by 10,000 samples. For comparison, the Penning ionization in the BEC for $\nu = 42$ would yield an averaged potential of -0.0099 eV.

In order to estimate the averaged potential, we use the Monte Carlo method as described in the previous subsection assuming that a pair of Rydberg atoms in the neighboring lattice sites with their overlapping Rydberg orbitals undergoes the Penning ionization. The ion positions are determined as follows: For size $m$ cluster, we randomly choose $\lfloor m/2 \rfloor$ Rydberg atoms in the cluster and replace them with the ions.

The Monte Carlo results are shown in Fig. S8. We show the averaged potential of the MI case as a function of the Rydberg excitation probability $p$. We can find that the averaged potential for $p = 0.03$ is larger than the electron binding energy for $\nu \geq 50$. The binding energy is obtained by using the open source program [48]. The potential due to the ions is thus estimated to be large enough to trap the Penning electrons.

In the BEC case for $\nu = 42$ and $p = 0.03$, we find that the averaged potential is about $U_{\text{ave}} = -0.0099$ eV. This value is comparable to the electron binding energy for $\nu = 42$: $E_{\text{bind}} = -0.0082$ eV. The potential may thus not be deep enough to trap the Penning electrons.